\documentclass[preprint]{aastex}
\usepackage[T1]{fontenc}
\usepackage[latin1]{inputenc}
\usepackage{graphics}
\begin{document}

\title{Modeling Fluctuations in the GRB-Afterglow Light
Curves}

\author{E. Nakar \& T. Piran}
\affil{Racah Institute for Physics, The Hebrew University,
Jerusalem 91904, ISRAEL}

\providecommand{\LyX}{L\kern-.1667em\lower.25em\hbox{Y}\kern-.125emX\@}


\begin{abstract}

The fluctuations observed in the light curves of some
GRB-afterglows (such as GRB 021004) provide a useful tool to probe
the circum-burst density profile  and to probe the variations in
the energy of blast-wave with time. We present a general formalism
that reduces the calculation of the observed light curve from a
Blandford-Mackee blast-wave to the evaluation of a one
dimensional integral. Using this formalism we obtain a simple
approximation to the general light curve that arises in more
complex situations where the afterglow's energy or the external
density vary. The solution is valid for spherically symmetric
profiles and it takes a full consideration of the angular time
delay effects. We present the light curves of several external
density profiles and demonstrate the effects of density
variations on the light curve. We also re-visit the afterglow of
GRB021004 and we find that the steep decay after the first bump
($\sim 4000sec$) cannot result from a spherically symmetric
density variation or from the passage of the synchrotron frequency
through the optical band. This suggests that an angular structure
is responsible to some of the observed features in the light
curve. This may be the first evidence that an angular structure
is important in the early stages of the afterglow.
\end{abstract}

\section{Introduction}

The basic theory of the multi wavelength afterglow which follows
GRBs is well established. The afterglow emission is produced by
an interaction between a relativistic expanding fireball and the
circumburst material. This interaction produces a relativistic
blast wave which heats the external medium and produces the
observed emission. The dominant radiation process is most likely
synchrotron.

After a short radiative phase the blast wave becomes adiabatic and
the hydrodynamic profile behind the shock relaxes to a
self-similar solution (Blanford \& Mackee 1976 ; hereafter BM76).
The observed synchrotron light curve from a Blanford-Mackee (BM)
self similar shell is presented by Granot, Piran \& Sari (1998) as
a two dimensional nontrivial numerical integral. We use here the
self similar properties of the BM solution to reduce the
expression of Granot et. al. (1998) to a simple one dimensional
integral. The integrand expresses the contribution of the
instantaneous emission from a BM shell at a given
radius\footnote{ The integrand includes an integration over the
angular and the radial dimensions which reduces, due to the
self-similar structure of the shell, to a general integral that is
calculated only once. }. This simplification is important for two
reasons. First, it allows an easy calculation of interesting
physical quantities (e.g. the afterglow image, see Granot et. al.
1998) that had to be calculated numerically so far. Second, and
more importantly, it enables us to obtain a good approximation of
the observed afterglow when the external density and/or
blast-wave energy varies, as long as the variations are
spherically symmetric.

We use our solution to investigate several external density
profiles. We find that due to angular smoothing, even a short
length scale density variation, results in a fluctuation in the
light curve of more than one order of magnitude in time. Second,
we show that above the cooling frequency, a density enhancement
results in a fluctuation with amplitude of less than 20\% , while
a density drop results in a larger fluctuation (a drop of one
order of magnitude in the density results in a fluctuation of
40\%).

Nakar, Piran \& Granot (2002) investigated different possible
explanations to the peculiar afterglow of GRB 021004. This
unusual afterglow shows a clear deviations from a smooth temporal
power law decay. A first bump is observed after \( \sim 4000sec
\), this bump is followed by a steep decay. Later the afterglow
shows additional deviations from a power-law decay. Nakar et al.
2002 considered the angular effects only approximately. Here we
re-visit this afterglow taking a full consideration of the
angular effects. We show that the first bump and the steep decay
that follows it, cannot result from spherically symmetric density
variations. We show also that the fast decay is inconsistent with
a simple transition of the typical synchrotron frequency,
$\nu_m$, through the optical band (with a constant ISM density
profile). Although passage of $\nu_m$ is consistent with the
timing and with the rising phase of the bump (Kobayashi \& Zhang
2002), the fast decay requires another mechanism (most likely an
angular dependent one). These results suggests that an angular
structure is important in the early afterglow of GRB 021004.

In \S2 we present our solution. In \S3 we consider the effects of
density and energy variations. In \S4 we discuss the afterglow of
GRB 021004 and at the last section we present our conclusions. The
detailed calculations which lead to the solution we present at \S2
are described in the appendix.

\section{The Blandford-McKee light curve}

We consider an adiabatic relativistic blast wave propagating into
an external medium with a regular density profile \(
n_{ext}\propto r^{-k} \) (\( k<3 \)). We calculate the observed
flux from a slow cooling synchrotron. As commonly done (Sari,
Piran \& Narayan 1998) we assume that the energy of the magnetic
field and the energy of the hot electrons, are constant fractions
of the internal energy (\( \varepsilon _{B} \) and \( \varepsilon
_{e} \) respectively) and that the hot electrons' initial
distribution is a power law with an energy index \( p \). We
express the solution as a simple one dimensional integral. This
solution can be used to calculate different properties of the
afterglow, or as an approximation to the observed afterglow in
the cases of a variable external density and/or a variable
blast-wave energy (see section \ref{sec e and n variations}).

The spectral shape of the observed flux is well approximated
(Sari et al. 1998) by a broken power law with several power-law
segments. In the slow cooling regime there are four segments
separated by \( \nu _{a},\nu _{m} \) and \( \nu _{c} \): the
self-absorption, synchrotron and cooling frequencies
respectively. We discuss here only the emission above \( \nu _{a}
\). First we consider the solution when the observed frequency,
\( \nu  \), is far from the break frequencies and later we
address the solution near the break frequencies.

We distinguish between three relative frames. The center of the
explosion and the observer at infinity are at rest at the source
frame. We denote the time in the source frame by \( t \) and we
calibrate \( t=0 \) to the explosion time. The fluid frame is the
rest frame of the fluid, we denote all variables in the fluid
frame with a 'prime'. The observer frame is at rest compared to
the source frame. However, the observer time is measured
according to the arrival time of photons to the observer. We
denote the observer time as \( T \), and we calibrate \( T=0 \)
to the arrival time of a photon emitted at \( t=0 \), \( r=0 \)
(\( r \) is the distance from the explosion).

We address first the problem of an instantaneous emission from a
very thin shell located at a radius \( r \) and propagating with
a Lorentz factor \( \gamma  \) (this problem or related ones
where considered by Fenimore, Madras \& Nayakshin (1996), Kumar
\& Panaitescu 2000, Ioka \& Nakamura 2001, Ryde \& Petrosian 2002
and others). Due to the relativistic beaming the observer receives
mainly photons emitted up to an angle of \( \theta =1/\gamma  \)
relative to the line-of-sight. The observer time delay between
two photons emitted simultaneously at a radius \( r \), one on
the line-of-sight and the other from \( \theta =1/\gamma  \) is
\( T_{ang}=r/2c\gamma ^{2} \). Hence, an instantaneous emission in
the source frame is observed as a finite pulse with a duration \(
\sim T_{ang} \). The observed time of the first photon (emitted
on the line-of-sight) is \( T_{los}=t-r/c \), where \( t \) is
the time in which the shell radiates in the source frame. Thus an
instantaneous emission from a very thin shell produces a pulse
that is observed for \( T_{los}\leq T\leq T_{los}+T_{ang} \). The
emission that arrives at larger \( T \) is emitted at larger \(
\theta  \). Fenimore et al. (1996) calculated the shape of the
pulse for an intrinsic power law spectrum with a power law index
\( \beta  \): \( P_{\nu }\propto \nu ^{\beta } \). They find that
the observed flux at \( T>T_{los} \) is:\begin{equation} \label{eq
Thin shell} F_{\nu }(T)\propto \nu ^{\beta }\left(
1+\frac{T-T_{los}}{T_{ang}}\right) ^{-(2-\beta )}.
\end{equation}

The observed flux, at an observer time \( T \), from an arbitrary
spherically symmetric emitting region is given by Granot et. al.
(1998):\begin{equation} \label{eq Fnu1} F_{\nu
}(T)=\frac{1}{2D^{2}}\int _{0}^{\infty }dt\int _{0}^{\infty
}r^{2}dr\int _{-1}^{1}d(cos\theta )\frac{n'(r)P'_{\nu }(\nu
\Lambda ,r)}{\Lambda ^{2}}\delta (t-T-\frac{r\, \, cos\theta
}{c}),
\end{equation}
 where \( n' \) is the emitters density and \( P'_{\nu } \) is the
emitted spectral power per emitter, both are measured in the
fluid frame; \( \theta  \) is the angle relative to the line of
sight, and \( \Lambda ^{-1}=1/\gamma (1-v\, \, cos\theta /c) \)
(\( v \) is the emitting matter bulk velocity) is the blue-shift
factor. Below we calculate this expression for an instantaneous
emission from a shell with a Blandford-Mackee (BM) self-similar
profile. The BM self similar shell has a width of, \( \Delta
R\approx R/2(4-k)\Gamma ^{2} \), where \( R \) and \( \Gamma  \)
are the shock's radius and Lorentz factor respectively. BM76 show
that all the variables behind the relativistic blast-wave are
functions only of the dimensionless parameter, \( \chi
=[1+2(4-k)\Gamma ^{2}](1-r/ct) \). \( \chi =1 \) at the shock
front and it increases with the distance (down stream) from the
shock.

We will calculate \( F_{\nu }(T) \) by integrating over the
contributions from BM shells (Eq. \ref{eq Fnu general} below) at
different radii. To do so we calculate the contribution to the
observed flux at time \( T \) of the emission emitted by the blast
wave during the time that the shock propagates from the radius \(
R \) to \( R+dR \). We denote this contribution as \( A_{\nu }(R)
\cdot g_{\beta }(R,T,k)dR \) where \( \beta  \) is the spectral
index in the relevant power-law segment. \( A_{\nu }(R)dR \) is
the emitted spectral power during the time that the shell
propagates from \( R \) to \( R+dR \).  The function, $ A_{\nu }$,
includes only numerical parameters which remain after the
integration over $r$ and $cos\theta$ in Eq. \ref{eq Fnu1} and it
depends only on the conditions of the shock front along the
line-of-sight. The second factor \( g_{\beta } \) is a
dimensionless factor that describes the observed pulse shape of an
instantaneous emission. \( g_{\beta } \) is obtained by
integration over \( cos\theta  \) and \( r \) in Eq. \ref{eq
Fnu1} and it includes only the radial and angular structure of
the shell. The self-similar profile of the shell enables us to
express \( g_{\beta } \) as a general function that depends only
on the dimensionless parameter \(
\widetilde{T}=\widetilde{T}(R,T) \) (which we define later),
hence \( g_{\beta }(R,T,k)=g_{\beta }(\widetilde{T},k) \). The
function \( g_{\beta }(\widetilde{T},k) \) is calculated only
once for a given \( {\beta } \) and external density profile.

Using the self similar profile of the emitting region and
neglecting terms up to the lowest order of \( O(\gamma ^{-2}) \)
(the detailed derivation is described in details in the Appendix),
we reduce Eq. \ref{eq Fnu1} to: \begin{equation} \label{eq Fnu
general} F_{\nu }(T)=\frac{1}{D^{2}}\int _{0}^{R_{max}}A_{\nu
}(R)g_{\beta }(\widetilde{T},k)dR,
\end{equation}
 where \( R_{max} \) satisfies\footnote{%
We denote values at the shock front as functions of the shock
radius alone (e.g. \( T_{los}(R)\equiv T_{los}(t(R),R) \). }
:\begin{equation} \label{eq Rmax} T_{los}(R_{max})=T,
\end{equation}
 and \( D \) is the distance to the source (cosmological factors are
not considered through the paper). When all the significant
emission from the shell at radius \( R \) is within the same
power-law segment, \( \beta  \), (i.e \( \nu  \) is far from the
break frequencies) then \( A_{\nu } \) and \( g_{\beta } \) are
given by: \begin{equation} \label{eq Anu general} A_{\nu
}(R)=H_{\nu }\left\{ \begin{array}{c} R^{2}\, n^{4/3}_{ext,0}\,
E_{52}^{1/3}\, M_{29}^{-1/3}\quad \nu <\nu
_{m}\\
R^{2}\, n^{(5+p)/4}_{ext,0}\, E_{52}^{p}\, M_{29}^{-p}\quad \nu
_{m}<\nu <\nu _{c}\\
R\, n^{(2+p)/4}_{ext,0}\, E_{52}^{p}\, M_{29}^{-p}\quad \nu >\nu
_{c}
\end{array}\right. \frac{erg}{sec\cdot cm\cdot Hz},
\end{equation}
where \( R \) is the radius of the shock front, \( n_{ext}(R) \)
is the external density, \( E \) is energy in the blast-wave, \(
M(R) \) the total collected mass up to radius \( R \) and \(
H_{\nu } \) is a numerical factor which depends on the observed
power law segment (see Eq. \ref{eq Hnu}). We denote by \( Q_{x} \)
as the value of the quantity \( Q \) in units of \( 10^{x} \)
(c.g.s).

\begin{equation}
\label{eq general pulse} g(\widetilde{T},\beta ,k)=\left\{
\begin{array}{c} \frac{2}{(4-k)}\int
^{1+2(4-k)\widetilde{T}}_{1}\chi ^{-\mu (\beta ,k)}\left(
1-\frac{1}{2(4-k)}+\frac{2(4-k)\widetilde{T}+1}{2(4-k)\chi
}\right) ^{-(2-\beta )}d\chi \quad \nu <\nu _{c}\\
(1+\widetilde{T})^{-(2-\beta )}\quad \nu >\nu _{c}
\end{array}\right. ,
\end{equation}
 where\begin{equation}
\label{eq Ttilda}
\widetilde{T}(R,T)=\frac{(T-T_{los}(R))}{T_{ang}(R)},
\end{equation}
 and\begin{equation}
\label{eq mu} \mu (\beta ,k)\equiv 3\cdot (71-17k)/(72-18k)-\beta
\cdot (37+k)/(24-6k).
\end{equation}
 This set of equations is completed with the following relations
between the different variables of the blast wave, the observer
time and the break frequencies:\begin{equation} \label{eq nu_{m}}
\nu _{m}=5\cdot
10^{12}n^{1/2}_{ext,0}E^{2}_{52}M^{-2}_{29}\varepsilon
^{1/2}_{B-2}\varepsilon _{e-1}^{2}\: Hz,
\end{equation}
\begin{equation}
\label{eq nu_{c}} \nu _{c}=2.5\cdot
10^{17}R_{17}^{-2}n_{ext,0}^{-3/2}\varepsilon ^{-3/2}_{B-2}\: Hz,
\end{equation}
\begin{equation}
\label{eq M(R)} M(R)\equiv 4\pi m_{p}\int ^{R}_{0}r^{2}n(r)dr\:
\left( =\frac{4\pi n_{*}m_{p}}{3-k}R^{3-k}\right) ,
\end{equation}
\begin{equation}
\label{eq Tlos(R)} T_{los}(R)=\frac{c}{4}\int
^{R}_{0}\frac{M(r)}{E(r)}dr\: \left( =\frac{\pi
cn_{*}m_{p}}{(4-k)(3-k)E}R^{4-k}\right) ,
\end{equation}
\begin{equation}
\label{eq Tang(R)} T_{ang}(R)=\frac{cRM(R)}{2E(R)}\: \left(
=2(4-k)T_{los}(R)\right) ,
\end{equation}
 where \( m_{p} \) is the proton mass and the values in the
parenthesis are for a constant energy and a density profile of \(
n_{ext}=n_{*}r^{-k} \) (\( k<3 \)). The numerical coefficient of
Eq. \ref{eq Anu general}, \( H_{\nu } \), is given
by:\begin{equation} \label{eq Hnu} H_{\nu =}\left\{
\begin{array}{c} 9\cdot 10^{-26}\, \varepsilon ^{2/3}_{e-1}\,
\varepsilon
^{1/3}_{B-2}\, \nu^{1/3}\quad \nu <\nu _{m}\\
2\cdot 10^{-21}(5\cdot 10^{12})^{(p-1)/2}\, \varepsilon
^{p-1}_{e-1}\, \varepsilon ^{(p+1)/4}_{B-2}\, \nu^{(1-p)/2}\quad
\nu _{m}<\nu <\nu _{c}\\
\frac{3p+6}{3p+2}6\cdot 10^{-3}(2.5\cdot 10^6)^p\, \varepsilon
^{p-1}_{e-1}\, \varepsilon ^{(p-2)/4}_{B-2}\, \nu^{-p/2}\quad
\nu >\nu _{c}
\end{array}\right. .
\end{equation}
For the two canonical cases of ISM (\( n_{ext}=n_{*} \), \( k=0 \))
and wind (\( n_{ext}=n_{*}r^{-k} \); \( k=2 \)) Eq. \ref{eq Anu
general} is reduced to:\begin{equation} \label{eq Anu ISM} A_{\nu
,los}(R)=H_{\nu }\left\{ \begin{array}{c}
2\cdot 10^{34}R_{17}\, n_{*,0}\, E_{52}^{1/3}\quad \nu <\nu _{m}\\
(0.07)^{-p}\cdot 10^{34}R_{17}^{2-3p}\, n^{(5-3p)/4}_{*,0}\,
E_{52}^{p}\quad \nu _{m}<\nu <\nu _{c}\\
(0.07)^{-p}\cdot 10^{17}R_{17}^{1-3p}\, n^{(2-3p)/4}_{*,0}\,
E_{52}^{p}\quad \nu >\nu _{c}
\end{array}\right. \frac{erg}{sec\cdot cm\cdot Hz}\; ISM,
\end{equation}
\begin{equation}
\label{eq Anu WIND} A_{\nu ,los}(R)=H_{\nu }\left\{
\begin{array}{c} 2\cdot 10^{34}R^{5/3}_{17}\, n_{*,34}\,
E_{52}^{1/3}\quad \nu <\nu
_{m}\\
(0.2)^{-p}\cdot 10^{34}R_{17}^{2-p}\, n^{(5-3p)/4}_{*,34}\,
E_{52}^{p}\quad \nu _{m}<\nu <\nu _{c}\\
(0.2)^{-p}\cdot 10^{17}R_{17}^{1-p}\, n^{(2-3p)/4}_{*,34}\,
E_{52}^{p}\quad \nu >\nu _{c}
\end{array}\right. \frac{erg}{sec\cdot cm\cdot Hz}\; WIND,
\end{equation}

For \( {\nu }>{\nu }_{c} \) the expression of \( g_{\beta
}(\widetilde{T},k) \) in Eq. \ref{eq general pulse} is analytic.
We can obtain an approximated analytic expression also for the
spectral segment \( \nu _{m}<\nu <\nu _{c} \). The rise time of
the pulse is very short (\( g_{\beta} \) reaches half of its
maximal value at \( \widetilde{T}\approx 0.02 \), and its maximal
value at \( \widetilde{T}\approx 0.14 \)). Hence, for \(
\widetilde{T}<0.25 \) we can approximate \( g_{\beta} \) as a
constant. For \( \widetilde{T}>0.25 \) the width of the BM shell
is negligible and we can approximate the pulse decay   as an
emission from a thin shell with an effective angular time, \(
T_{ang}^{eff} \). We calculate \( T_{ang}^{eff} \) by
approximating the BM profile as a series of thin shells whose
Lorentz factors that vary with \( \chi  \), emitting at the same
time an radius\footnote{ This approximation is not valid for \(
\nu <\nu _{m} \). In this spectral range the contribution of
shells with large \( \chi  \) decay more slowly, and the width of
the BM shell at late times, but not to late (\(
\widetilde{T}\approx 1 \)) can not be neglected. } . Due to the
different Lorentz factors  each shell has its own angular time
(\( T_{ang}(\chi )\propto \chi \)). The effective angular time is
a weighted average of \( T_{ang}(\chi ) \). The weights are the
emitted spectral power density at \( \chi  \): \( \propto \chi
^{-\mu (\beta ,k)} \) (see the appendix for details). Hence: of the
\begin{equation}
\label{eq g approx} g_{\beta }(\widetilde{T},k)\approx \left\{
\begin{array}{c}
g_{\beta }(0.25,k)\quad \widetilde{T}<0.25\\
g_{\beta }(0.25,k)\left( \frac{1+\widetilde{T}(\mu -2)/(\mu
-1)}{1+(\mu -2)/4(\mu -1)}\right) ^{-(2-\beta )}\quad
\widetilde{T}>0.25
\end{array}\right. \nu _{m}<\nu <\nu _{c},
\end{equation}
\begin{equation}
\label{eq Tang effective} T^{eff}_{ang}=\frac{\mu -1}{\mu
-2}T_{ang}(R)\quad \nu _{m}<\nu <\nu _{c}.
\end{equation}

The validity of this approximation is shown in Fig \ref{Fig BM
pulse shape}, which compares the approximation of Eq. \ref{eq g
approx} with the complete calculation of Eq. \ref{eq general
pulse} and a full numerical simulation of the emission from a BM
blast wave. The numerical simulations include both the adiabatic
and the radiative cooling of the electrons.

\begin{figure}
\resizebox*{0.9\columnwidth}{0.3\textheight}{\includegraphics{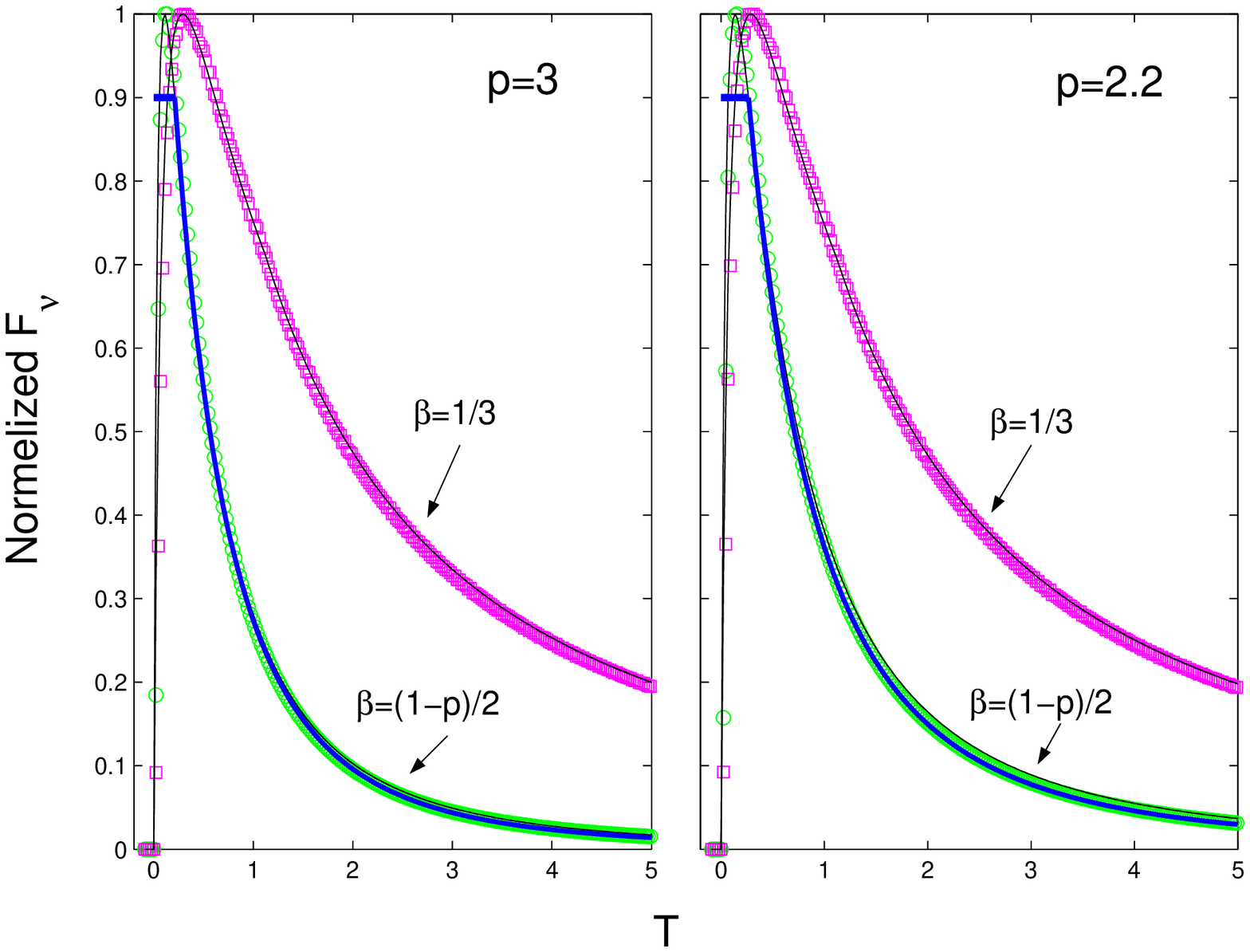}

} \caption{\label{Fig BM pulse shape}.The general pulse shape in
ISM, $ g_{\beta }(\widetilde{T},k=0) $, for electrons' energy
indices $p=2.2$ and $p=3$. The thin solid lines corresponds to Eq.
\ref{eq general pulse}, the thick lines are the results of Eq.
\ref{eq g approx}, and the squares and the circles are the results
of a full numerical simulation of the emission from a BM blast
wave. Note that the approximation of Eq. \ref{eq g approx} is very
good during the decay of the pulse. It is almost impossible to
distinguish between the approximation and the exact calculation of
$g_{\beta }(\widetilde{T},k=0)$ at this phase. At
$\widetilde{T}>10$, the approximation is less accurate, but at
this stage the contribution of the pulse to the observed light
curve is negligible. Similar results are obtained for other $k$
values. }
\end{figure}

\subsection{The light curve in the vicinity of the break frequencies}

The above solution is valid only when the observed frequency is
far from any of the break frequencies (\( \nu _{m} \) and \( \nu
_{c} \)). To understand the behavior in the vicinity of the break
frequencies we consider a thin shell with an intrinsic
broken power-law spectrum. \( \nu '_{b} \) is the break frequency
in the shell's rest frame (\( \beta =\beta _{1} \) for \( \nu
'<\nu '_{b} \) and \( \beta =\beta _{2} \) for \( \nu '>\nu '_{b}
\)). \( \nu '_{b} \) is constant along the shell; at the observer
frame, however, the break frequency is \( \nu _{b}=\nu
'_{b}/\Lambda  \) (\( \Lambda ^{-1} \) is the blue-shift factor
which defined after Eq. \ref{eq Fnu1}), which decrease with \(
\theta  \). Hence, it is possible that at first the observed
frequency, \( \nu \), is smaller than \( \nu _{b} \), while at
later times it is larger than \( \nu _{b} \). In other words at
different times the observed frequency is within different
power-law segments. In this case Eq. \ref{eq Thin shell} is still
valid, only \( \beta =\beta _{1} \) for \( \nu /\nu
_{b}>(1+(T-T_{los})/T_{ang}) \) and \( \beta =\beta _{2} \) for
\( \nu /\nu _{b}<(1+(T-T_{los})/T_{ang}) \).

When the instantaneous emission is from a BM shell then the
blue-shift varies within different parts of the shell (larger \(
\chi  \) and/or larger \( \theta  \) result in a smaller \(
\Lambda ^{-1} \)). Hence, the observed flux from an instantaneous
BM shell at a given observer frequency at a given time results
from emission in a range of fluid frame frequencies. Therefore,
it is possible that the flux in a given observed frequency at a
given time corresponds to different power-law segments at the
emission from different parts of the shell.

For \( \nu \ll \nu _{c} \) the observed emission arrives from the
whole width of the shell. The integration over the parameter \(
\chi  \) within the pulse shape, \( g \), is both along the
radial coordinate and the angular coordinate (larger \( \chi  \)
is smaller \( r \) and lower \( \theta  \)). This integration
depends on the spectral index \( \beta  \), which is different
for \( \nu <\nu _{m} \) (\( \beta _{1}=1/3 \)) and \( \nu >\nu
_{m} \) (\( \beta _{2}=(1-p)/2 \)). The transition from one
spectral index to the other occur at some critical \( \chi  \)
value, \( \chi _{m} \), which satisfies:\begin{equation}
\label{eq ChiM } \chi _{m}^{\xi }(1-\frac{1}{2(4-k)})+\chi
_{m}^{\xi -1}(\widetilde{T}+\frac{1}{2(4-k)})-\frac{\nu _{m}}{\nu
}=0
\end{equation}
 where \( \xi =(37+k)/(24-6k) \). Here the pulse shape depends also
on \( \chi _{m} \) and it takes the form: \begin{eqnarray} A_{\nu
}g_{\beta _{1},\beta _{2}}(\widetilde{T},k,\chi
_{m})=\frac{2}{(4-k)}\left[ A_{\nu <\nu _{m}}\int ^{\chi
_{m}}_{1}\chi ^{-\mu (\beta _{1},k)}\left( 1-\frac{1-{\chi
}^{-1}}{2(4-k)}+\frac{\widetilde{T}}{\chi }\right) ^{-(2-\beta
_{1})}d\chi \right.  &  & \nonumber \\
\left. +A_{\nu >\nu _{m}}\int ^{1+2(4-k)\widetilde{T}}_{\chi
_{m}}\chi ^{-\mu (\beta _{2},k)}\left( 1-\frac{1-{\chi
}^{-1}}{2(4-k)}+\frac{\widetilde{T}}{\chi }\right) ^{-(2-\beta
_{2})}d\chi \right] \quad \nu <\nu _{c}\label{eq g ChiM}
\end{eqnarray}
The weight of the contribution of each power-law segment is given
by the corresponding \( A_{\nu } \). Whenever \( \chi _{m}<1 \)
than the whole shell emits within the same power-law segment, \(
\nu >\nu _{m} \), and Eq. \ref{eq g ChiM} is reduced to Eq.
\ref{eq general pulse}. Similarly, when \( \chi
_{m}>1+2(4-k)\widetilde{T} \) the whole shell emits within the
power-law segment of \( \nu <\nu _{m} \). Eq. \ref{eq g ChiM}
provides an exact solution of the spectral break at \( \nu _{m} \)
and an exact light curve break when \( \nu _{m} \) passes through
the observed frequency. Just like Eq. \ref{eq general pulse}, \(
g_{\beta _{1},\beta _{2}}(\widetilde{T},k,\chi _{m}) \) is
calculated only once. However in this case it should be calculated
for every \( \widetilde{T} \) and \( \chi _{m} \).

When \( \nu \approx \nu _{c} \) the emission arrives only from a
thin part at the front of the BM shell. The local spectrum of the
emission in the fluid frame vary along the shell, and an exact
solution should follow the exact profile of the local emission.
Hence, there is no simple solution for the exact flux in the
vicinity of \( \nu _{c} \) (a full solution of the break shape in
ISM and wind is presented at Granot \& Sari 2001). A partial
treatment of the break is obtained by taking a sharp transition
from \( \nu <\nu _{c} \) to \( \nu >\nu _{c} \) when \( \nu
(1+\widetilde{T})=\nu _{c} \), i.e taking the part of \( \nu <\nu
_{c} \) at Eq. \ref{eq general pulse} for \( \widetilde{T}<\nu
_{c}/\nu -1 \) and the part of \( \nu >\nu _{c} \) for \(
\widetilde{T}>\nu _{c}/\nu -1 \). In this approximation \(
g_{\beta _{2},\beta _{3}}(\widetilde{T},k) \) is discontinuous
(\( \beta _{3}=-p/2 \)), but the observed light curve and spectral
break are rather smooth.

\subsection{The emission from a collimated jet}

So far we have dealt with a spherical symmetric systems. However,
in GRBs the relativistic outflow is most likely collimated into
narrow jets with an opening angle \( \theta _{j} \). Our solution
is not valid if the hydrodynamical parameters depend on the angle
from the jet axis. But if they do not, then we can easily
generalize our results to a jet as long as the observation angle
relative to the jet axis is much smaller than \( \theta _{j} \).
In this case the emission from the edges of the jet at a given \(
R \) is observed at \( \widetilde{T_{j}}(R,T)=(\gamma (R)\theta
_{j})^{2} \) as long as \( \widetilde{T_{j}}\geq 1 \). Hence, in
this case \( g_{\beta }\left( \widetilde{T}(R,T)>(\gamma (R)\theta
_{j})^{2},k\right) =0 \) and we can use all the above equations
with this substitution. The hydrodynamic evolution of such a jet
is similar to spherical symmetric evolution as long as \( R \)
satisfies \( \widetilde{T_{j}}=(\gamma (R)\theta _{j})^{2}>1 \).
For larger radii the hydrodynamical evolution changes (the jet
spreads sideway) and a jet break is observed in the light curve.
The effect of the cutoff, \( \widetilde{T_{j}} \), on the observed
light curve is negligible for \( \nu >\nu _{m} \) and the
spherical symmetric solution is valid for any observed time before
the jet break time. For \( \nu <\nu _{m} \) the decay is slower,
and taking \( g_{\beta }=0 \) at the edges of the jet is required
also before the break. Clearly the whole solution is not valid
after the jet break.

\section{Density and Energy variations\label{sec e and n variations}}

In the previous section we have calculated the observed light
curve, for a regular external density and a constant energy
blast-wave. Consider now the effect of variations in the external
density or in the energy of the blast-wave. If the variations are
not too rapid, then the shell profile behind the shock can be
approximated by a BM self-similar profile with the instantaneous
energy and external density. The light curve can be expressed as
an integral over the emission from a series of instantaneous BM
solutions.

It is worthwhile to explore the conditions in which this
approximation is valid. We consider first density variations and
then we turn to energy variations. When a blast wave at radius \(
R \) propagates into the circumburst medium, the emitting matter
behind the shock is replenished within \( \Delta R\approx
R(2^{1/(4-k)}-1) \). This is the length scale over which an
external density variation relaxes to the BM solution. Our
approximation is valid as long as the density variations are on a
larger length scales than \( \Delta R \) . Our approximation is
not valid when there is a sharp density increase over a range of
\( \Delta R \). However, the contribution to the integral from
the region on which the solution breaks is small (\( \Delta
R/R\ll 1 \)) and the overall light curve approximation is
acceptable. Note, however, that a density jump by more than a
factor of \( \sim 21 \) can produce a reverse shock (Dai \& Lu
2002) which breaks the BM profile of the shell and the validity
of our approximation.

A sharp density decrease is more complicated. Here the length
scale in which the emitting matter behind the shock is
replenished could be of the order of \( R \). As an example we
consider a sharp drop at some radius \( R_{d} \) and a constant
density for \( R>R_{d} \). In this case the external density is
negligible at first, and the hot shell cools by adiabatic
expansion. Later the forward shock becomes dominant again. Kumar
and Panaitescu (2000) show that immediately after the drop the
light curve is dominated by the emission during the adiabatic
cooling. Later the the observed flux is dominated by emission
from \( R\approx R_{d} \), and at the end the new forward shock
becomes dominant. Our approximation includes the emission before
the density drop and the new forward shock after the drop, but it
ignores the emission during the adiabatic cooling phase.

A sharp density drop with \( n_{ext}\propto r^{-k} \) with \( k>4
\) or an exponential drop which continues over a long length
scale breaks the BM solution and therefore our approximation
breaks down. Some of these cases can be described by self-similar
solutions (Best \& Sari 2000 ; Perna \& Vietri 2002; Wang, Loeb
\& Waxman 2002). We do not consider these cases here. However our
calculations can be followed with the new self-similar profiles.

\begin{figure}
\resizebox*{0.9\textwidth}{0.3\textheight}{\includegraphics{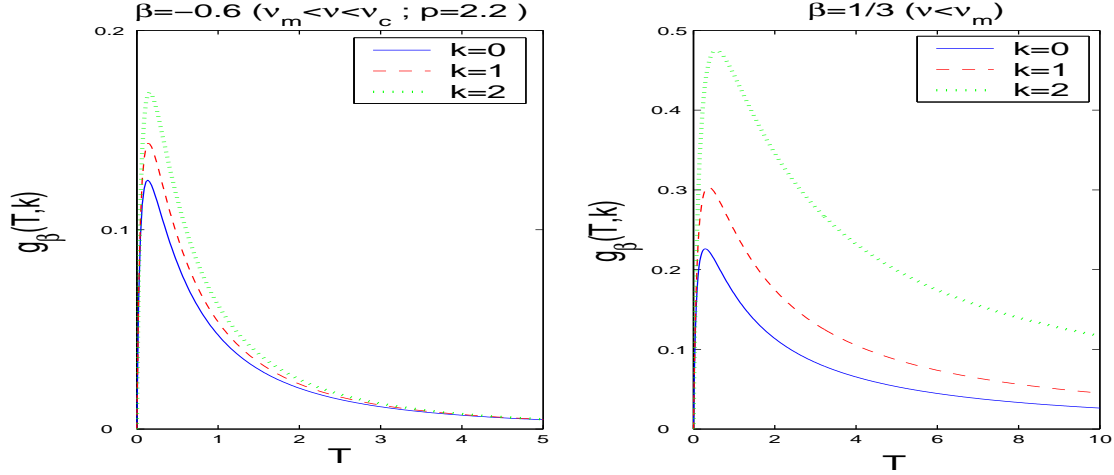}}
\caption{\label{Fig k dependance}$g_{\beta}(\widetilde{T},k)$ for
different $k$ values. Left) $\nu_m < \nu < \nu_c$; The
differences between the shape of the pulses are small. The rise
and the decay times are similar for different $k$ values, and the
the peaks' heights ratios between different $k$ pulses are less
then 1.5.  Right) $\nu < \nu_m$; The differences between the
shape of the pulses are large, mainly in the decay time.
Therefore, at this spectral segment it is important to consider
the changes in $k$ when the density varies.}
\end{figure}

An additional effect of density variations arises from the
relation \( k\equiv -(R/n_{ext}(R))(dn_{ext}/dR) \). When the
density varies so does \( k \). This effect is important only at
\( \nu <\nu _{m} \). For \( \nu >\nu _{c} \), the light curve does
not depend on \( k \). Fig \ref{Fig k dependance}a depicts \(
g_{\beta }(\widetilde{T},k) \) when \( \nu _{m}<\nu <\nu _{c} \)
for different \( k \) values. It shows that \( g_{\beta } \)
depends weakly on \( k \) in this spectral segment. Fig \ref{Fig k
dependance}b shows that for \( \nu <\nu _{m} \) a re-calculation of
\( k \) with \( R \) is needed.

Spherically symmetric energy variations are most likely to occur
due to refreshed shocks, when new inner shells arrive from the
source and refresh the blast wave ( Rees \& Meszaros 1998, Kumar
\& Piran 2000, Sari \& Meszaros 2000). Kumar \& Piran (2000) show
that in such case the solution has a smooth transition from the
BM solution with the energy of the pre-collision blast-wave ( the
front shell) to another BM solution with the total energy of the
two shells. The collision produces, however, a reverse shock
whose emission has a lower peak frequency than the forward shock
emission. Clearly our approximation fails to capture the effect
of the reverse shock and it does not capture the details of the
light curve during the time that the shock crosses the outer
shell.

\begin{figure}
\resizebox*{0.9\textwidth}{0.3\textheight}{\includegraphics{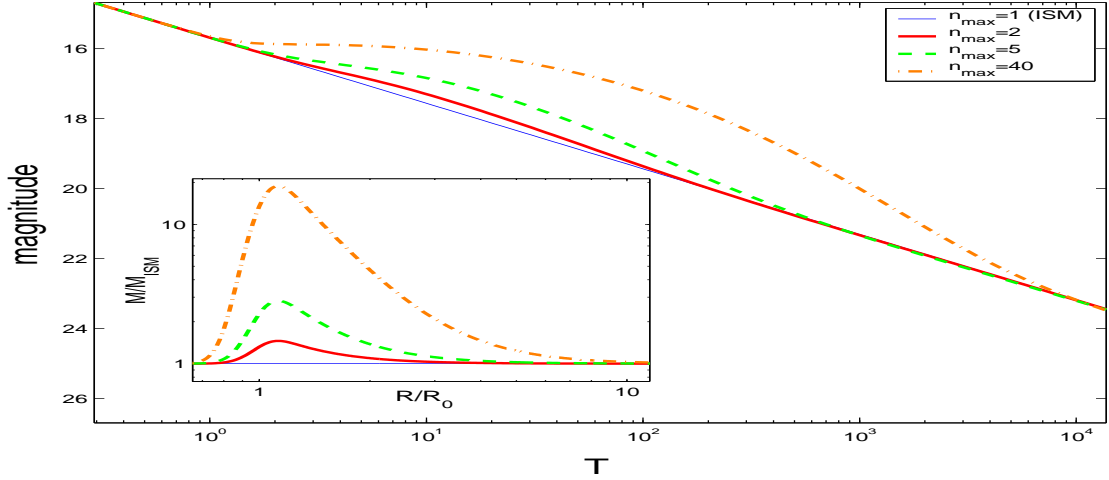}}
\caption{\label{Fig Gaussian over-density} The light curves results from a
Gaussian ($\Delta R/R=0.1$) over-dense region in the ISM. The
different thick lines are for a maximal over-densities of $40$
({\it dashed-dot}), $5$ ({\it dashed}) and $2$ ({\it solid}). The
thin line is the light curve for a constant ISM density. The
inset depicts the ratio of the mass, $M(R)$ over the mass of an
ISM (without the over-dense region), $M_{ISM}(R)$.}
\end{figure}

\begin{figure}
\resizebox*{0.4\textwidth}{0.3\textheight}{\includegraphics{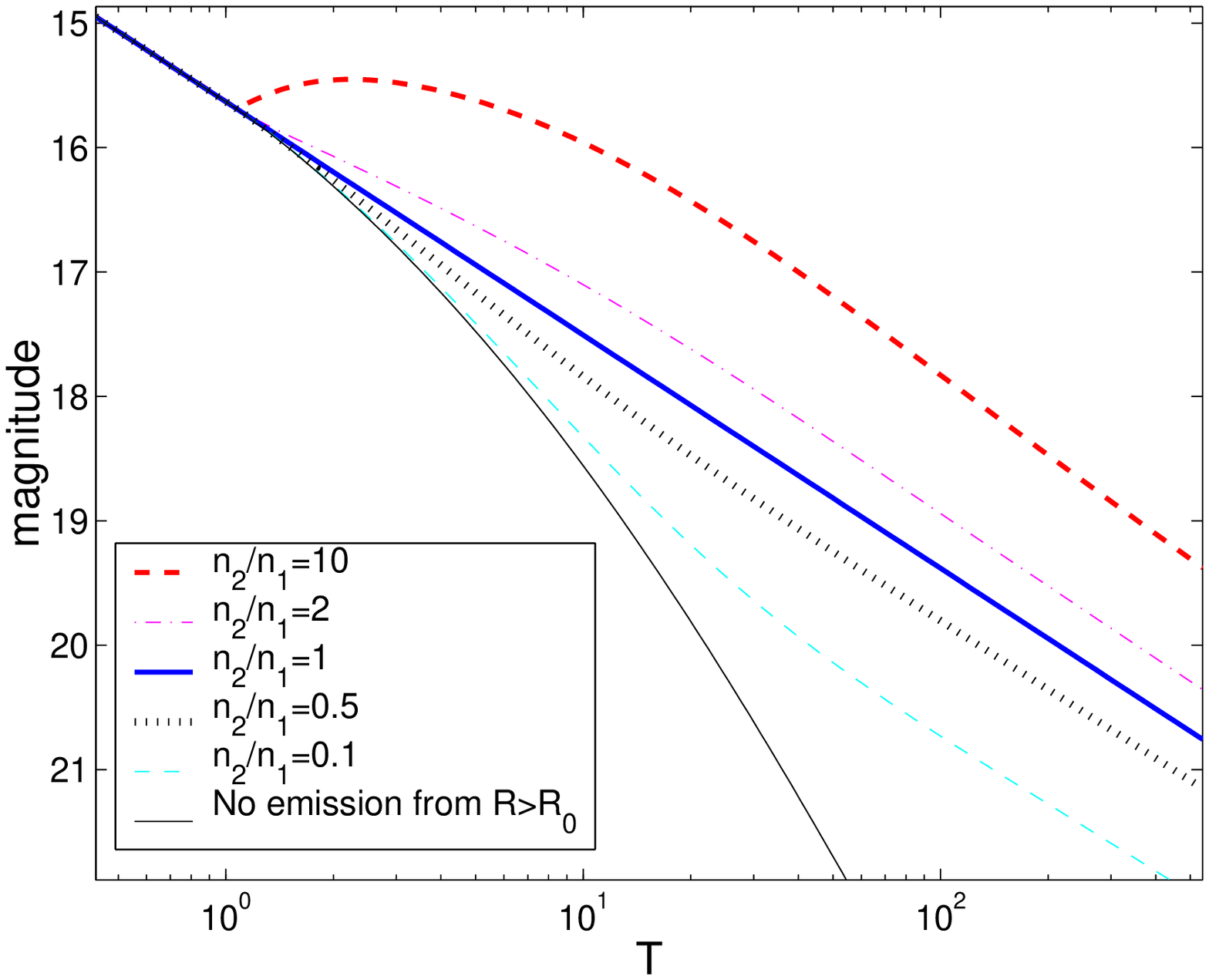}}
\resizebox*{0.4\textwidth}{0.3\textheight}{\includegraphics{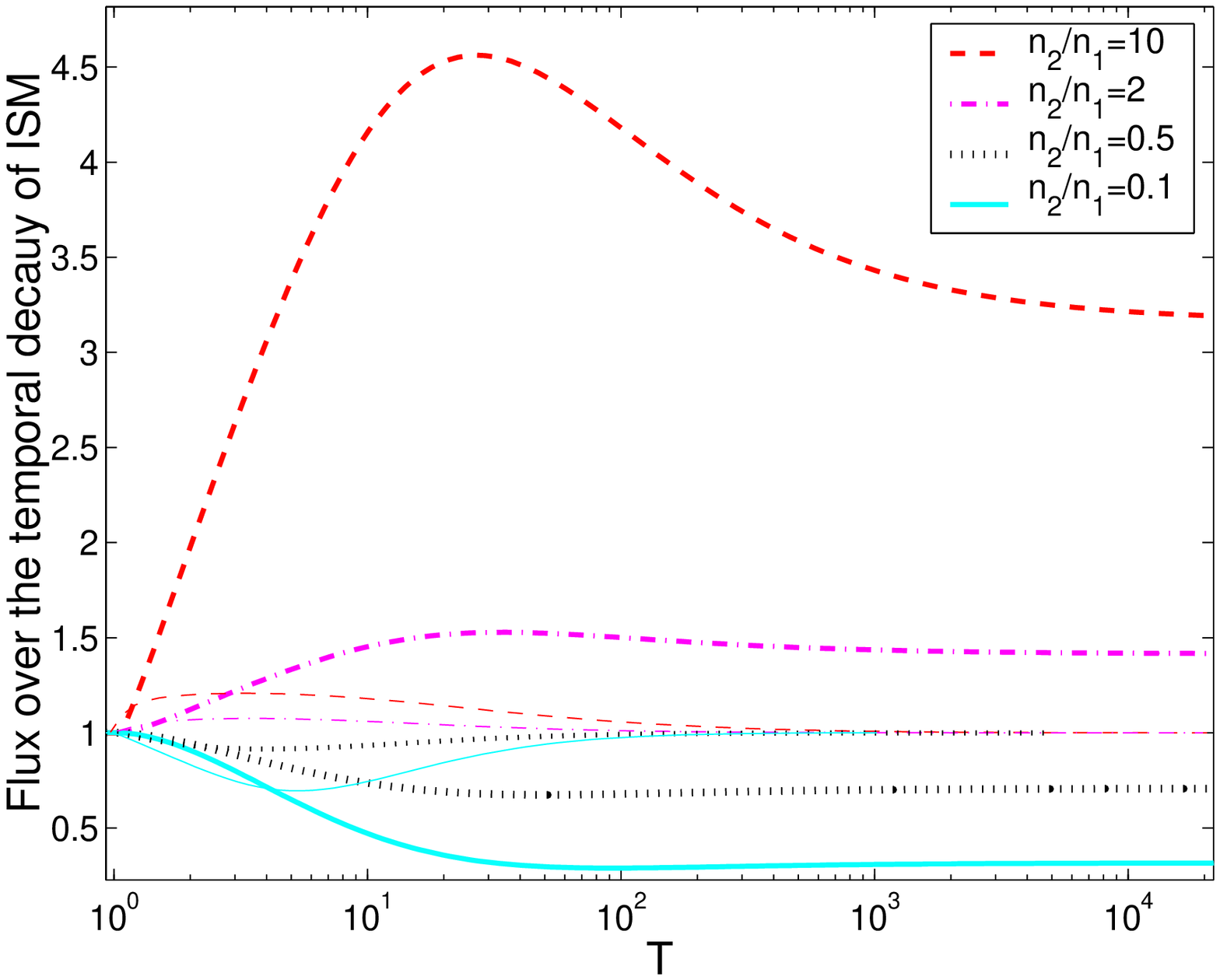}}
\caption{\label{Fig desnity jump} Left) The light curves for
several density jumps and drops ($n=n_{1}$, $R<R_{0}$; $
n=n_{2}$, $R>R_{0}$) for $\nu _{m}<\nu <\nu _{c}$. Right)  The
flux deviation from an ISM light curve produced by the same
density jumps and drops for $ \nu _{m}<\nu <\nu _{c}$ (\emph{bold
lines}) and $\nu _{c}<\nu$ (\emph{thin lines}). In both panels
the time, $T$, is normalized so $T_{los}(R_0)=1$.}
\end{figure}

Our method enables a simple calculation of the observed light
curve for a given density and energy profile. In Figures \ref{Fig
Gaussian over-density} and \ref{Fig desnity jump} we show the
observed light curves for several different density profiles (with
constant energy). Fig \ref{Fig Gaussian over-density} depicts the
\( \nu _{m}<\nu <\nu _{c} \) light curve for a Gaussian ($\Delta
R/R=0.1$) over-dense region in the ISM. Such a density profile
may occur in a clumpy environment. The emission from a clump is
similar to the emission from a spherically over-dense region as
long as the clump's angular size is much larger than \( 1/\gamma
\). The different light curves are for a different maximal
over-densities. We find that a maximal over-density of \( 5 \)
effects the observed light curve during two orders of magnitude
in time. The effect of a maximal over-density of 40 is observed
during four order of magnitude in time (note that in all the cases
the width of the Gaussian is similar). Even a mild short
length-scale, over-dense region (with a maximal over-density of
2)  influences the light curve for a long duration (mainly due to
the angular effects). This duration depends strongly on the
magnitude of the over-density. Note that due to the nonlinear
dependence of the observed flux on $n_{ext}$ a narrower Gaussian
(smaller $\Delta R/R$) with an equivalent amount of mass in the
over-dense region produces a larger fluctuation.

Fig \ref{Fig desnity jump} shows the observed light curve for
several density jumps and drops. In the left panel \( \nu
_{m}<\nu <\nu _{c} \) at all times. In the right panel we compare
the effect of similar density jumps on the light curve above and
below \( \nu _{c} \). When \( \nu _{m}<\nu <\nu _{c} \) there is
a transition from one power-law to another with the same power
law index and a flux ratio factor of \( (n_{2}/n_{1})^{1/2} \)
between the two power-laws  (as expected according to Sari et al.
1998). The duration of the transition is longer for larger density
contrasts. This transition is observable for a duration of about
two orders of magnitude in time for a density contrast of \( 10
\) and for one order of magnitude in time for a density contrast
of \( 2 \). When \( \nu _{c}<\nu  \) (right panel) then after a
small fluctuation, the light curve returns to behave as if there
was no jump (or drop). The width of this fluctuation is between
three orders of magnitude in time for high density contrast (\(
10 \)) to two orders of magnitude in time for small density
contrast (2). The maximal amplitude obtained for a density jump
is \( \sim 20\% \) (a deviation of \( \sim 0.2mag \)) while the
deviation for a density drop can reach \( \sim 40\% \) (a
deviation of \( \sim 0.4mag \))

Now, using Eqs. \ref{eq Fnu general}-\ref{eq Tang(R)} we can
approximate the observed light curve for given energy and density
profiles. However, for a given burst we usually have the observed
light curve at hand and not the energy and density profiles at
the source. In this case we can invert Eqs. \ref{eq Fnu
general}-\ref{eq Tang(R)} (numerically) under the assumption of
either a constant energy or a constant density. Thus, we can find
the profile of the free variable, which produces
the observed light curve\footnote{%
This procedure is similar to the one we have used in Nakar et al.
(2002), however here we take a complete consideration of the
angular effects. }. The analytic approximation of \( g_{\beta }
\) (Eq. \ref{eq g approx}) greatly simplifies this inversion when
\( \nu _{m}<\nu  \). The observed light curve at a given time is
a convolution of emission at many different source times (or
shock front's radii). Inverting Eqs. \ref{eq Fnu general}-\ref{eq
Tang(R)} requires a de-convolution of the light curve to the
emission at different radii. Unfortunately, deconvolution
amplifies small errors in the observed data and the resulting
de-convolved signal (or in our case the energy or density profile)
is highly sensitive to small variations in the observed light
curve.

In some cases the inversion of the observed light curve fails.
This usually happens when the light curve depicts rapid decay.
The angular and radial spreading dictates a fastest possible
temporal decay (see Eqs. \ref{eq general pulse} and \ref{eq g
approx} in which \( F_{\nu }\propto T^{-(2-\beta )} \) at late
times). A faster temporal decay is impossible even if the emission
from the blast wave completely stops. A faster observed temporal
decay would result in a failure to invert Eqs. \ref{eq Fnu
general}-\ref{eq Tang(R)}. This failure implies that a new effect (like
angular dependence), which we do not consider, must be included.

\section{GRB 021004}

The peculiar afterglow of GRB 021004 was observed on October 4'th
2002. The early optical detection (Fox et al. 2002), \( T \sim 500sec
\), enabled a detailed observation of this afterglow from a very
early stage. This unusual afterglow shows clear deviations from
a smooth temporal power law decay. A first bump is observed at
\(T \sim 4000sec \), this bump is followed by a steep decay.
Another smaller bump is observed at \(T \sim 7\cdot 10^{4}sec \)
and a possible third one at \( 3\cdot 10^{5}sec \). A steepening
which may be a jet break is observed at \( \sim 7days \). Several
different mechanisms can lead to these observations. Some of the
machisms suggested so far are external density variations, angular
energy structure (patchy shell model), refreshed shocks, and a passage
of \( \nu _{m} \) through the optical band (Lazzati et al. 2002,
Kobayashi \& Zhang 2002, Nakar et al. 2002, Holland et al. 2002,
Pandey et al. 2002, Bersier et al. 2002, Schaefer et al. 2002, Heyl
\& Perna 2002). The last scenario (the passage of \( \nu _{m} \))
explains only the first bump, and is combined with the emission of
the reverse shock which should be dominant till the first bump.

\begin{figure}
\resizebox*{0.9\textwidth}{0.3\textheight}{\includegraphics{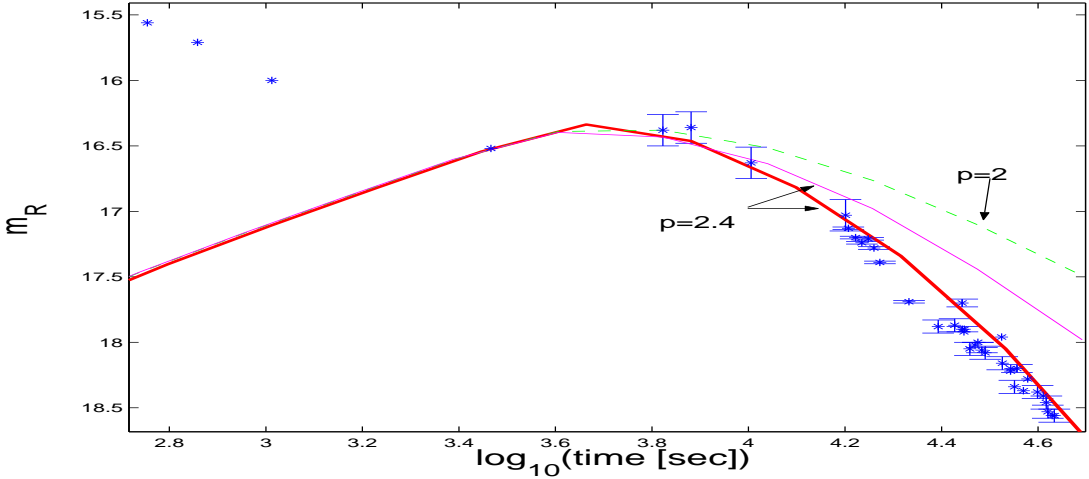}
} \caption{\label{Fig nu m passage}The best fit to the first bump
in the R-band observations of GRB 021004 (see the references to
the GCNs at Nakar et al. 2002), using the passage of $\nu_m$
through the optical band in an ISM density profile. The different
thin lines are for $p=2$ ({\it dashed}) and $p=2.4$ ({\it solid}).
The thick line is for $p=2.4$, assuming that the emission is
completely stopped just when \( \nu _{m} \) is in the \( R \)
band.}
\end{figure}

In the following we apply our method to two possibilities\footnote{%
Our spherically symmetric model is not applicable to the patchy
shell model and refreshed shocks can not explain the fluctuations
below the expected power-law decay. } (i): A spherically symmetric
(or an angular scale larger then \( 1/\gamma  \)) density
variations and constant energy; assuming that the optical emission
is at all time above \( \nu _{m} \) and below \( \nu _{c} \), (ii)
The passage of \( \nu _{m} \) through the optical band, assuming
an ISM density profile and a constant energy. Figure \ref{Fig
varying density} depicts the best fit that we obtained with a
spherically symmetric density variations with \( p=2 \). In order
to get the fastest decay possible after the bump, we stop the
emission completely at the peak of the bump. It is clear that due
to the angular spreading, it is impossible to fit the fast decay
after the first bump. The fit is even worse with larger \( p \) .
Figure \ref{Fig nu m passage} depicts the best fit of the first
bump for a passage of \( \nu _{m} \) through the R band in an ISM
density profile and with a constant energy. Again, due to the
angular spreading, it is impossible to fit the fast decay. We
obtain a marginally consistent fast decay only if we assume that
the emission is completely stopped just when \( \nu _{m} \) is in
the R-band. But clearly such a coincidence is unlikely.

The conclusion from these results is that it is unlikely that the
light curve of GRB 021004 results from a spherically symmetric
fluctuations. This result provides a new evidence that also in the
early times of the GRB emission, an angular structure (either of
the relativistic wind or the circum-burst medium) is involved.

\begin{figure}
\resizebox*{0.9\textwidth}{0.3\textheight}{\includegraphics{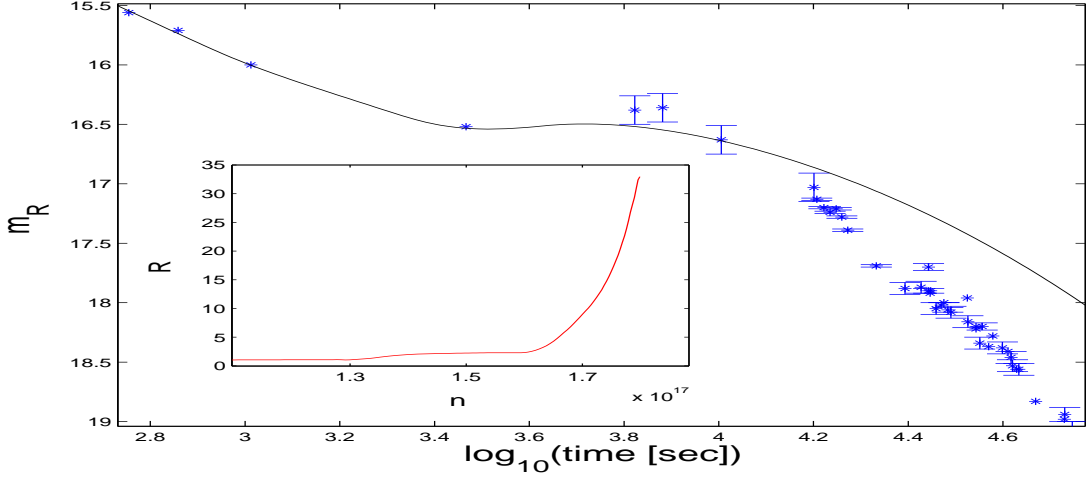}
} \caption{\label{Fig varying density} The best fit to the R-band
observations of GRB 021004 (see the references to the GCNs at
Nakar et al. 2002), using a varying density (and constant energy).
We assume that $\nu_m<\nu<\nu_c$ at all times and we take $p=2$,
$E=10^{52} \rm ergs$, $ \varepsilon _B=0.01$ and $\varepsilon
_e=0.1$. The inset depicts the density profile which produces this
light curve. Note that at $R \sim 1.8 \cdot 10^{17}$ The density
drops to zero (the emission stops). Even though, the resulting
light curve cannot fit the steep decay of the observations.}
\end{figure}

\section{Conclusions}

We have presented a simplified solution of the slow cooling
synchrotron emission form a BM blast wave. This solution separates
the observed flux at a given time to the contributions from
different BM shells with a different radius of the shock front.
Using the self-similar profile of the BM shells we have shown that
the pulse shape of the emission from BM shells at different radii
is general (independent of the shock's front radius). We have also
presented an analytic expression to this pulse shape for $\nu >
\nu_m$. Thus, this pulse shape could be calculated only once (or
the analytic expression may be used), and the whole solution turns
into a simple one dimensional integral over the contributions from
different radii. This simplification enables an easy calculation
of different properties of the afterglow which until now had to be
calculate using a complicate and computational time consuming
simulations.

The main advantage of our solution is that it enables us to
approximate the emission from a blast-wave with a varying energy
and/or a varying external density, as long as these variations are
spherically symmetric. The advantage of this solution over the
method we have presented in Nakar et al. (2002) is the full
consideration of the angular effects.

We use our solution to approximate the light curve which results
from several density profiles. We find out that the duration of
fluctuations in the light curve, which results from density
variations, are long even if the length scale of the density
variation, $\Delta R$, is very short ($\Delta R/R \ll 1$). For
example a density variation with $\Delta R/R = 0.1$ and a mild
over-density results in a fluctuation which is observed for two
orders of magnitudes in time. We show also that density variations
induce mild ($\sim 30\%$) fluctuations also above $\nu_c$.
Fluctuation induced by a density drop are larger than the
fluctuation induced by a density jump. These fluctuations are
also observed for about two orders of magnitude in time in the
case of a sharp density jump, or drop.

We try to fit the early afterglow of GRB 021004, by a spherically
symmetric varying density and by the passage of $\nu_m$ through
the optical band. Both fits fail to follow the fast decay after
the first bump in the afterglow. This results suggests that an
angular structure within the ejecta or within the external density
is crucial for the production of the early afterglow of GRB
021004.

\section*{Appendix: Derivation of the Light Curve Formula\label{sec full calc}}

In the appendix we show the details of the calculations which
lead to Eqs. \ref{eq Fnu general}-\ref{eq Hnu}. First we solve
the problem for \( \nu <\nu _{c} \) and later we consider the
solution for \( \nu >\nu _{c} \).

We start from Eq. \ref{eq Fnu1} which gives the observed flux
from an arbitrary spherically symmetric emitting region. For
$cos{\theta}=c(t-T)/r$ we obtain:\begin{equation} \label{eq Doppler}
\Lambda (r,t)=\gamma (1-v\, \, cos\theta /c)\approx (2\gamma
)^{-1}\left( 1+\frac{T-T_{los}(r,t)}{T_{ang}(r,t)}\right) .
\end{equation}
 At a given \( t \), the front of the blast wave is at radius \( R(t)
\) and the emitting region is restricted to \( r<R(t) \). On the
other hand, emission from \( r<r_{min}=c(t-T)=R(t)-c(T-T_{los}(R))
\) would not reach the observer at time \( T \). Hence,
integrating over $cos{\theta}$ {[}\( \delta (cos{\theta} )=c\delta (t-T-rcos{\theta}
/c)/r \){]}, and keeping only terms of the lowest order of \(
\gamma ^{-2} \), Eq. \ref{eq Fnu1} is reduced to:\begin{equation}
\label{eq Fnu2} F_{\nu }(T)\approx \frac{2}{D^{2}}\int
_{0}^{R_{max}}dR\int _{r_{min}}^{R}drr\gamma ^{2}n'(r)P'_{\nu
'm}\left( \frac{\nu }{2\gamma \nu '_{m}}\right) ^{\beta }\left(
1+\frac{T-T_{los}(r)}{T_{ang}(r)}\right) ^{-(2-\beta )},
\end{equation}
 where \( R_{max} \), is the maximal radius of the shock, from which
an emission from the blast wave contributes to the flux at time
\( T \), i.e. \( T_{los}(R_{max})=T \) . We also used the
spectrum for \( \nu <\nu _{c} \): \( P'_{\nu }(\nu \Lambda
,r,t)=P'_{\nu ',m}(r,t)\cdot (\nu \Lambda /\nu '_{m})^{\beta } \),
where \( \nu '_{m} \) is the synchrotron frequency in the fluid
frame and \( \beta  \) is the spectral index.

The dependence of the hydrodynamical parameters in a BM self
similar shell on \( \chi  \) is (BM76): the bulk Lorentz factor, \(
\gamma (\chi )=\Gamma \sqrt{1/2\chi }=\gamma (R)/\sqrt{\chi } \),
the internal energy density in the fluid rest frame, \( e'(\chi
)=e'(R)\cdot \chi ^{-(17-4k)/(12-3k)} \) and the fluid density
behind the shock in the observer frame, \( n(\chi )=n'(\chi
)\gamma (\chi )=n(R)\cdot \chi ^{-(7-2k)/(4-k)} \), where \(
\gamma (R) \), \( e(R) \) and \( n(R) \) are the hydrodynamical
parameters values at the shock front (at radius \( R \)). Now, We
can express the observed synchrotron frequency, \( \nu
_{m}=2\gamma \nu '_{m}(\chi ) \), and the observed spectral power
at this frequency \( P_{\nu ,m}=\gamma P'_{\nu ',m} \) (Sari et
al. 1998) as a function of \( \chi  \): \( \nu _{m}(\chi )\propto
\gamma B\gamma _{m}^{2}\propto \gamma (\chi )\sqrt{e(\chi
)}\gamma _{m}^{2}(\chi ) \) and \( P_{\nu ,m}(\chi )\propto
\gamma (\chi )B\propto \gamma (\chi )\sqrt{e(\chi )} \), where \(
\gamma _{m} \) is the minimal Lorentz factor of the hot electrons
distribution. In the slow cooling regime the radiative cooling of
the electrons is negligible. The adiabatic cooling of a single
electron is proportional to \( e'(\chi )/n'(\chi ) \), hence \(
\gamma _{m}(\chi )=\gamma _{m}(R)\cdot \chi ^{-(2+2k)/(12-3k)} \).
Now, we can represent all the variables in Eq. \ref{eq Fnu2} as a
function of the shock front, \( R \), and the dimensionless
parameter \( \chi \) (which increase with the distance from the
shock front). Integrating over \( \chi  \) {[}\( dr=-d\chi
R/2(4-k)\Gamma ^{2} \){]} , using the relation:
\begin{equation}
\label{eq T-Tlos} \frac{T-T_{los}(r(\chi ))}{T_{ang}(r(\chi
))}=\frac{\widetilde{T}(R)}{\chi }-\frac{\chi -1}{2(4-k)\chi },
\end{equation}
 (\( \widetilde{T}(R) \) is defined in Eq. \ref{eq Ttilda}) and
expressing the density behind the shock as \(
n(R)=2n_{ext}(R)\Gamma ^{2} \) (see BM76) we obtain:
\begin{equation} \label{eq Fnu nu<nuc} F_{\nu }(T)\approx
\frac{2}{(4-k)D^{2}}\int _{0}^{R_{max}}A_{\nu }(R)g_{\beta
}(\widetilde{T}(R),k)dR\quad \nu <\nu _{c},
\end{equation}
 where \begin{equation}
\label{eq Anu<nuc} A_{\nu }(R)=R^{2}n_{ext}(R)P_{\nu ,m}(R)\left(
\frac{\nu }{\nu _{m}(R)}\right) ^{\beta }\quad \nu <\nu _{c}
\end{equation}
 and \( g_{\beta }(\widetilde{T}(R),k) \) is defined by Eq. \ref{eq
general pulse}. Note that the emitted spectral power density
depends on \( \chi  \) as: \( \gamma ^{2}n'P'_{\nu 'm}(\gamma \nu
'_{m})^{-\beta }\propto \chi ^{-\mu (\beta ,k)} \) (\( \mu  \) is
defined in Eq. \ref{eq mu}). The values of \( \gamma (R) \), \(
\nu _{m}(R) \) and \( P_{\nu ,m}(R) \) could be found for any
given external density profile (see Sari et al. 1998 and Nakar et
al. 2002) as we have done in Eq. \ref{eq Anu general}.

For \( \nu >\nu _{c} \) the emitting electrons are cooling fast,
and only a very thin layer behind the shock contributes to the
emission at this spectral regime. Therefore the pulse shape of an
instantaneous emission from a blast wave at radius \( R \), \(
g_{\beta }(\widetilde{T},k) \), is similar to the pulse shape of
an instantaneous emission from a very thin shell (see Eq. \ref{eq
Thin shell}) with a spectral index of \( \beta =-p/2 \). Finelly,
the emitted spectral power at the shock front, \( A_{\nu }(R) \),
is \begin{equation} \label{eq Anu>nuc} A_{\nu
}(R)=R^{2}n_{ext}P_{\nu ,m}\nu _{m}^{-\beta }\nu ^{1/2}_{c}\nu
^{\beta -1/2}\quad \nu >\nu _{c},
\end{equation}
 where \( \beta  \) is the spectral index for \( \nu _{m}<\nu <\nu
_{c} \).

\end{document}